\documentclass[twocolumn]{aastex702}
\usepackage{amsmath}

\usepackage{xcolor}
\definecolor{skyblue}{rgb}{0.53, 0.81, 0.92}
\definecolor{deepskyblue}{rgb}{0.0, 0.75, 1.0}

\newcommand{\jades}{JADES}
\shorttitle{FM-JADES-v1}
\shortauthors{Ding et al.}

\begin{document}

\title{Learning JWST. I. A Foundation Model for New Population Discoveries and Morphology-Aware Photometric Redshift Measurements in the JADES Survey}


\author[0000-0003-4651-8510]{Jiani Ding$^{*}$}
\email{jianiding@arizona.edu}
\affiliation{Steward Observatory, University of Arizona, 933 North Cherry Ave., Tucson, AZ 85721, USA}

\author[0000-0002-5367-8021]{Minghao Yue$^{*}$}
\email{yuemh@arizona.edu}
\affiliation{Steward Observatory, University of Arizona, 933 North Cherry Ave., Tucson, AZ 85721, USA}

\author[0000-0003-3307-7525]{Yongda Zhu}
\email{}
\affiliation{Steward Observatory, University of Arizona, 933 North Cherry Ave., Tucson, AZ 85721, USA}

\author[0000-0003-3310-0131]{Xiaohui Fan}
\email{}
\affiliation{Steward Observatory, University of Arizona, 933 North Cherry Ave., Tucson, AZ 85721, USA}

\author[0000-0002-4623-0683]{Yufeng Luo}
\email{}
\affiliation{Department of Physics and Astronomy, University of Wyoming, Laramie, WY 82071, USA}

\begingroup
\renewcommand{\thefootnote}{*}
\footnotetext{These authors contributed equally to this work.}
\endgroup


\begin{abstract}
We present FM-JADES-v1, a self-supervised foundation model for James Webb Space Telescope ({\em JWST}) deep-field science,
trained with 482,444 objects from the {\em JWST} Advanced Deep Extragalactic Survey (JADES) Data Release 5
using multi-band imaging and the photometric catalog. The shared embedding space is trained without class labels. We demonstrate that FM-JADES-v1 can serve as a powerful tool for object discovery and improving property measurements using two experiments, blind active discovery and few-band photometric redshift. For blind object discovery,
FM-JADES-v1 identifies rare object populations such as high-redshift galaxies and Little Red Dots (LRDs) without any prior population labels or population-specific selection criteria. These rare populations emerge as isolated islands in the embedding space, which can be identified without prior astrophysical knowledge.
For few-band photometric redshift, FM-JADES-v1's learned embeddings achieve $\sigma_{\rm NMAD}=0.157$ 
in a strictly controlled three-band (F115W/F200W/F356W) photo-$z$ benchmark, compared to $\sigma_{\rm NMAD}= 0.44$ for template fitting. These results demonstrate the potential of self-supervised multi-modal representations as scalable discovery spaces for large astronomical surveys. Applied to ongoing and future wide-field surveys from JWST, Roman, Euclid, and Rubin/LSST, this framework could enable systematic searches for rare populations, as well as enabling multiple downstream tasks such as improving astrophysical property measurements.

\end{abstract}


\section{Introduction} \label{sec:intro}

The {\em James Webb Space Telescope (JWST)} has revolutionized our understanding of distant galaxies by providing unprecedentedly accurate measurements and uncovering previously unknown or underexplored populations.
However, most {\em JWST} discoveries are based on predefined color cuts, spectral diagnostics, or serendipitous findings during visual inspection. This is particularly true for rare objects such as high-redshift galaxies \citep[e.g.,][]{Finkelstein2022,Harikane2023,Curtislake2023,ArrabalHaro2023,Bunker2023,Castellano2024,Carniani2024,naidu26} and Little Red Dots (LRDs) \citep[e.g.,][]{Matthee2024, Greene2024,Kokorev2024,Kocevski2024,akins25}. Such passive, selection-driven approaches are highly biased and becoming increasingly impractical for ongoing and upcoming large surveys like the Dark Energy Spectroscopic Instrument \citep[DESI; e.g.,][]{Dey2019,DESI2023a,DESI2023b}, Euclid \citep{Euclidc2026}, the Vera C. Rubin Observatory Legacy Survey of Space and Time \citep[LSST;][]{lsst}, and the Nancy Grace Roman Space Telescope surveys \citep{Akeson2019,ROTAC2025}, which will deliver vastly larger imaging and spectroscopic datasets. Instead, a scalable discovery framework must be developed to enable active data-driven searches that can jointly learn from multiple modalities to systematically identify unusual objects and previously unrecognized populations without predefined selection criteria. 

Self-supervised learning (SSL) provides a promising approach for efficiently identifying valuable objects and improving physical property measurements in large survey datasets. 
Instead of optimizing directly for a single scientific task, SSL models 
 are trained through surrogate objectives \citep[like reconstructing masked inputs, distinguishing augmented views, and aligning different data modalities; e.g.,][]{chen2020,Grill2020,He2022} to learn compact embeddings that can retain a physically informative structure in the datasets, without requiring labeled training sets \citep[e.g.,][]{chen2020,Hayat2021,Stein2021,MohaleLochner2024}. When learned on a sufficient scale and designed to transfer across datasets and applications, SSLs form the basis of foundation models \citep{Bommsani2021}. Foundation models provide an efficient approach to encode complex multi-modal observations from sky surveys into a shared and scientifically informative representation space \citep{Hayat2021}. The common representation can be used in a wide range of downstream applications, including similarity search for known objects, classification, anomaly detection, and physical property measurements \citep[e.g.,][]{Parker2025}.

 Recent efforts have explored the application of foundation models to astronomical data, covering both imaging and spectroscopy surveys \citep{Zhang2024,2024Smith,Parker2025,Aayush2025,Shao2026}, including several downstream tasks such as classification and estimates of physical parameters \citep[][]{Parker2024, Zhao2025, Shen2026, Euclidfm2026}. However, the application of foundation models on {\em JWST} datasets, in particular the extragalactic surveys, have remained largely unexplored. It is unclear how to use SSL and foundation models to enhance data analysis and scientific discovery for {\em JWST} data and how to transfer the {\em JWST-learned} knowledge to future sky surveys.

Motivated by this, we initiated a series of papers entitled {\em learning JWST}, with the aim of exploring efficient deep learning methods in the {\em JWST} datasets, including legacy surveys like JADES (this paper) \citep[e.g.,][]{Eisenstein2026,Robertson2026} and NIRCam grism surveys (Paper II) \citep[e.g.,][]{Sun2025,Wang2023ASPIRE,Kashino2023EIGER}.
As the first of this series, this work presents FM-JADES-v1, a self-supervised cross-modal foundation model trained on the JADES DR5 data product \citep{Carreira2026,Eisenstein2026,Johnson2026,Robertson2026}. The model jointly leverages the DR5 multi-band JWST/NIRCam imaging and photometric catalog measurements to learn a shared representation that encodes both morphological and photometric information.
Using this model, we demonstrate how to use representative learning to guide the discovery of new objects. Astrophysical meaningful structures can emerge without population labels in this self-supervised multi-modal representation of the survey data, and candidate populations are identified and validated using the structure encoded.  We further illustrate that 
representations learned from multi-band photometry and imaging significantly improve photometric redshift estimates compared to template fitting when only a small number of bands are available. 
Together, these results highlight foundation models as powerful tools for active discoveries of previously-underexplored object populations, as well as for improving the measurements of object properties.

The remainder of this paper is organized as follows. In \S~2, we introduce the JADES data used in this paper and describe the associated pre-processing procedures. In \S~3, we present the detailed model architecture and the training results. \S~4 introduces an active blind search framework to discover rare or previously unrecognized populations, using high-redshift galaxies and LRDs as examples. In \S~5, we investigate morphology-aware photometric-redshift (photo-$z$) measurement and show how representations learned from multi-band photometry and morphology can improve photo-$z$ estimation when only a limited number of photometric bands are available. \S~6 and \S~7 are the discussion and conclusion.

\section{Data} \label{sec:data}

\subsection{JADES Data Release 5} \label{sec:trainingdata}

We use the \jades\ Data Release 5 (DR5)\footnote{https://jades-survey.github.io/scientists/data.html} 
NIRCam mosaic images and photometric catalogs in GOODS-S and GOODS-N fields \citep{Johnson2026,Robertson2026,Eisenstein2026}. The union
footprint covered by at least one NIRCam filter is 245\,arcmin$^2$ in GOODS-S
and 224\,arcmin$^2$ in GOODS-N (469\,arcmin$^2$ in total). The image input
uses the 16 filters available in both fields: the wide filters F070W, F090W,
F115W, F150W, F200W, F277W, F356W, and F444W, and the medium filters F162M,
F182M, F210M, F300M, F335M, F410M, F430M, and F460M. Together, they span
approximately 0.7--4.6 \,$\mu$m. 
The DR5 mosaics are heterogeneous rather than a single 16-band footprint: among these filters, the individual-filter
coverage ranges from 10.07 to 233.71\,arcmin$^2$ in GOODS-S and from 8.74 to
207.53\,arcmin$^2$ in GOODS-N. The corresponding median, aperture-corrected
$5\sigma$ point-source depths range from 28.28 to 30.07\,AB mag and from
27.78 to 29.61\,AB mag, respectively \citep{Robertson2026}.
The pixel scale of the mosaic images is $0\farcs03$.
JADES DR5 also contains F250M and F480M imaging in GOODS-S, but these
field-specific filters are not used by the image encoder.
This decision was made to prevent the model from learning which field the objects are located.

The JADES DR5 catalog contains 485,510 objects, providing their forced circular-aperture and elliptical Kron
photometry, morphology, and photometric-redshift measurements. 
The photometry measurements cover 35
filters: 18 {\em JWST}/NIRCam, nine {\em HST}/ACS or WFC3, and eight
{\em JWST}/MIRI filters. We use the subset of these products (described below) and
refer to \citet{Robertson2026} for catalog construction.
procedures, and full column definitions. 

\subsection{Training Data Preparation} \label{sec:dataprep}

To construct the imaging and catalog training set, we make $64\times64$ pixel cutouts around the center of each object in each band 
and exclude objects with no valid image pixels in their extracted stamps. We further exclude objects with fewer than two
valid bands from the photometric catalog. This leads to a training set of 482,444
objects, containing 302,471 in GOODS-S and 179,973 in GOODS-N. We store the source ID,
right ascension, and declination as metadata, although the metadata are not used as model inputs.


For each object, we standardize its multi-band image as follows. For object $i$, we denote its flux and uncertainty in band $b$ at pixel $p$ as $f_{ibp}$ and $\sigma_{ibp}$, and compute the global median of the object as

\begin{equation}
{\rm med}_i=\underset{(b,p)\in V_i}{\operatorname{median}}f_{ibp},
\label{eq:image_scale}
\end{equation}

where $V_i$ is the set of valid (i.e., unmasked) science pixels.
We then compute a scale factor $s_i$ as

\begin{equation} 
s_i=1.4826\,\underset{(b,p)\in V_i}{\operatorname{median}}
\left|f_{ibp}-
{\rm med}_i\right|\times\alpha
\label{eq:image_scale}
\end{equation}

where 1.4826 is the conversion factor between the Median Absolute Deviation (MAD) and the standard deviation of Gaussian distributions, and $\alpha$ is a tunable constant. 
We then compute the standardized science and error images of the object as 

\begin{equation}
x^{\rm sci}_{ibp}=\operatorname{asinh}(f_{ibp}/s_i),\qquad
x^{\rm err}_{ibp}=\operatorname{asinh}(\sigma_{ibp}/s_i),
\label{eq:image_transform}
\end{equation}

Here we use the asinh function to suppress the contribution from very bright pixels. For pixels with $f_{ibp}\gg s_i$, we have $x^{\rm sci}_{ibp}\sim\ln (f_{ibp}/s_i)$. This approach prevents the brightest pixels from dominating the training loss. The constant $\alpha$ in Equation \ref{eq:image_scale} controls the characteristic flux above which the flux response becomes roughly logarithmic. 
In this study, we use $\alpha=8$, meaning that signals lower than 8$\sigma$ levels remain approximately linear in this conversion, while higher signals have approximately logarithmic response. This value $\alpha$ is chosen by running a prototype model for 50,000 training objects, with the goal of maximizing the effective dimension for the resulting embedding vector of the sample (Section \ref{sec:architecture}).
We note that the prototype test returns similar results for $6\lesssim\alpha\lesssim14$.
The model reads $[x^{\rm sci}_{ibp}, x^{\rm err}_{ibp}, {M}_{ibp}]$ as input, where ${M}_{ibp}$ is the valid pixel mask.

For the photometric catalog, we choose the following columns as input.
\begin{itemize}
    \item Aperture photometry. The JADES catalog provides aperture photometry for 35 bands over seven aperture sizes. For each band and each aperture size, we obtain the flux and uncertainties (e.g., \texttt{F444W\_CIRC1}, \texttt{F444W\_CIRC1\_e}, and \texttt{F444W\_CIRC1\_ei}). We then convert the fluxes and errors using a signed log1p function, $g(y)=\operatorname{sign}(y)\log_{10}(1+|y|)$. We compute soft Signal-to-Noise Ratio (SNR) 
     as softSNR$={\rm asinh}({\rm flux}/{\rm error})$.
    \item Kron photometry. Same as aperture photometry, we obtain the Kron and Kron\_S fluxes, uncertainties, and soft SNRs. We also include \texttt{A\_KRON}, \texttt{B\_KRON}, \texttt{THETA\_KRON}, \texttt{A\_KRON\_S}, \texttt{B\_KRON\_S}, \texttt{THETA\_KRON\_S} columns.
    \item Morphology parameters. We obtain \texttt{A}, \texttt{B}, \texttt{FWHM}, \texttt{GINI}, \texttt{NPIX\_DET}, \texttt{R\_KRON\_U}, \texttt{BBOX\_XMIN}, \texttt{BBOX\_XMAX}, \texttt{BBOX\_YMIN}, \texttt{BBOX\_YMAX}, \texttt{THETA} columns from the JADES DR5 catalog SIZE extension.
    \item Photometric redshifts. We obtain the following columns from the PHOTOZ extension: \texttt{z\_a} (the fiducial photo-$z$), \texttt{z\_ml} (the maximum likelihood redshift), \texttt{z\_peak} (the redshift with maximum $\exp(-\chi^2)$), in addition to the following columns: \texttt{z025}, \texttt{z160}, \texttt{z500}, \texttt{z840}, \texttt{z975}, \texttt{l68}, \texttt{u68}, \texttt{l95}, \texttt{u95}, \texttt{l99}, \texttt{u99}, \texttt{nfilt}, \texttt{Prob\_gt\_[z]} where $\texttt{[z]}\in(5,6,7,8,9)$, and \texttt{Prob\_z\_bins}.
\end{itemize}

We further derive the following quantities from the columns listed above:

\begin{itemize}
    \item Colors. We compute 73 colors using all 35 {\em HST}+NIRCam+MIRI bands. Specifically, we include all 34 adjacent-filter pairs, all 33 skip-one-filter pairs, plus six broad-band colors: F606W-F814W, F814W-F115W, F150W-F200W, F277W-F356W, F356W-F444W, and F444W-F770W. All colors are computed using \texttt{CIRC0} magnitudes.
    
    \item Half-light radius index. For each object and each band, we linearly interpolate the circular fluxes with aperture 0 to 6 onto the aperture indices, and compute the index corresponding to $0.5\times\max(f_{\rm{circ},i})$, where $f_{\rm{circ},i}$ denotes all the aperture fluxes. We denote this index as \texttt{rhalf\_circ\_index}. 
\end{itemize}

We provide colors and half-light radius as input features so that the model does not have to relearn simple relations among the photometric measurements. This allows the representation to focus more on the higher-order structure in the data.

We refer the reader to \cite{Robertson2026} for the detailed meaning of each column. Before training, all columns are standardized using the sample mean and standard deviation. We also replace missing values with placeholders with explicit validity flags. The adopted catalog columns and their related tokens (\S \ref{sec:tokenization}) are listed in Table \ref{tab:catalog_tokens}.

\begin{deluxetable*}{
>{\raggedright\arraybackslash}p{0.2\textwidth}
>{\raggedright\arraybackslash}p{0.1\textwidth}
>{\raggedright\arraybackslash}p{0.4\textwidth}
>{\raggedright\arraybackslash}p{0.27\textwidth}}
\tablecaption{Catalog tokens and their related original columns
\label{tab:catalog_tokens}}
\tablehead{
\colhead{Token Name} & \colhead{Number of Tokens}  &  \colhead{Description} &
\colhead{Corresponding catalog columns} 
}
\startdata
\texttt{phot:[band]:[aperture]} & 315 & One token for each band and each aperture describing the flux and error. The bands include 18 NIRCam bands, nine {\em HST} bands, and eight MIRI bands. The apertures include circular aperture 0--6, as well as Kron and Kron\_s.
& \texttt{[band]\_[aperture]}, \texttt{[band]\_[aperture]\_e}, \texttt{[band]\_[aperture]\_ei}  from extension CIRC\_BSUB\\
\texttt{rhalf:[band]} & 35 & One token for each band describing the estimated half-light radius. See text for details.
& \texttt{rhalf\_circ\_index}, calculated using \texttt{[band]\_CIRCi} where $\texttt{i}\in$(0,1,2,...,6) from extension CIRC\_BSUB\\
\texttt{color:[band1]:[band2]} & 73 &
Colors between selected pairs of filters (see text for details), calculated using their \texttt{CIRC0} magnitudes &
\texttt{[band1]\_CIRC0}, \texttt{[band2]\_CIRC0}  from extension CIRC\_BSUB. \\
\texttt{morph:[key]} & 2 & Morphological parameters measured via Kron photometry, where \texttt{key}  $\in$ [\texttt{Kron}, \texttt{Kron\_s}] & 
\texttt{[A\_[key], B\_[key], theta\_[key]]} from catalog extension \texttt{KRON}\\
\texttt{morph:miri\_[key]} & 2 &
Same as \texttt{morph:[key]}, but for MIRI measurements &
\texttt{[A\_[key], B\_[key], theta\_[key]]} from catalog extension \texttt{MIRI} \\
\texttt{morph:general} & 1 &
Summary of the morphology information &
\texttt{A, B, FWHM, GINI, NPIX\_DET, R\_KRON\_U, BBOX\_XMIN, BBOX\_XMAX, BBOX\_YMIN, BBOX\_YMAX, THETA}  from extension SIZE \\
\texttt{photoz:summary} & 1 &
Summary of the EAZY \citep{Brammer2008} photometric fitting result &
\texttt{z\_a}, \texttt{z\_ml}, \texttt{z\_peak}, \texttt{z025}, \texttt{z160}, \texttt{z500}, \texttt{z840}, \texttt{z975}, \texttt{l68}, \texttt{u68}, \texttt{l95}, \texttt{u95}, \texttt{l99}, \texttt{u99}, \texttt{nfilt}, \texttt{Prob\_gt\_[z]} where $\texttt{z} \in(5,6,7,8,9)$, from PHOTOZ\_KRON extension\\
\texttt{photoz:posterior} & 1 &
 Posterior distribution of photometric redshift from EAZY &
\texttt{[prob\_z\_bin00, \ldots,
prob\_z\_bin21]}  from PHOTOZ\_KRON extension\\
\enddata
\tablecomments{All catalog columns also receive a valid flag, which is passed to the catalog tokenizer (\S\ref{sec:tokenization}) together with the catalog value. The model therefore receives the validity of all catalog values.}
\end{deluxetable*}


\section{Model} \label{sec:model}

\begin{figure*}
\centering
\includegraphics[width=0.95\textwidth]{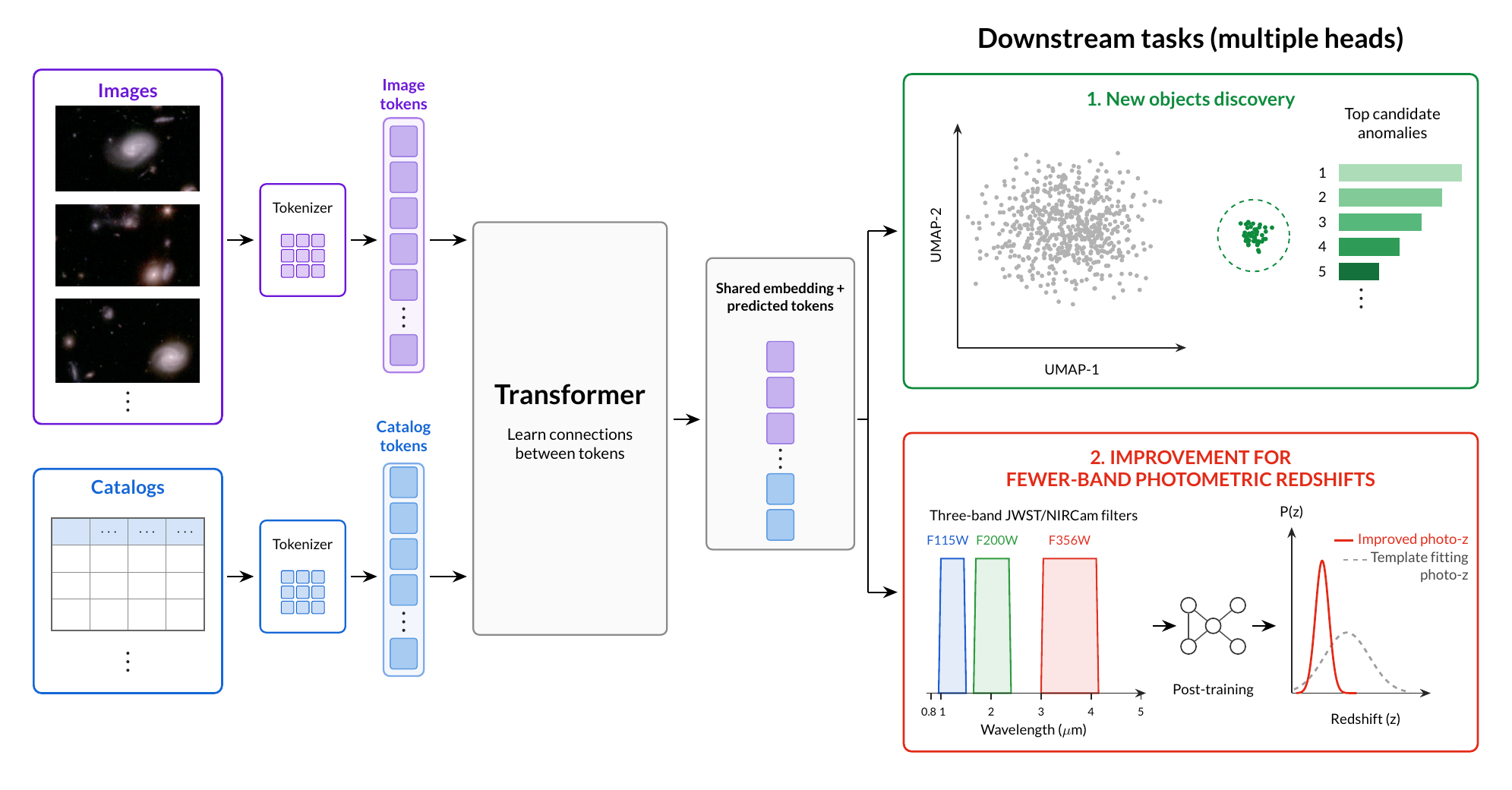}
\caption{Schematic overview of the \textsc{FM-JADES-v1} architecture and downstream tasks. Multi-band images and source catalogs are first tokenized separately into image tokens and catalog tokens. Then a multi-modal Transformer will jointly process the tokens to learn cross-token and cross-modal relationships. The model outputs a shared embedding together with predicted tokens for downstream tasks. The two main downstream tasks from this model are illustrated in the right panel: (1) New object discoveries using the shared embedding (Section \ref{sec:blind_discovery}); (2) Using image and catalog tokens to improve photo-$z$ measurements (Section \ref{sec:photozapp}). 
\label{fig:arch}}
\end{figure*}

Our model design is motivated by that of AION-1 \citep{Parker2025}. Figure~\ref{fig:arch} demonstrates the overall model architecture; in short, we first convert the standardized images and catalogs into a series of tokens, then use a transformer model to learn the internal relations between the tokens. 
We use a composite loss function to ensure that the model learns astrophysical information. Below we provide details about each component of the model. 

\subsection{Tokenization} \label{sec:tokenization}


In this model, all objects and data are described as a series of tokens. 
Specifically, we use the following form to tokenize an object and its associated data:

\[
[\mathrm{CLS}]\oplus[\mathrm{NORM}]\oplus
[\mathrm{IMAGE}]\oplus[\mathrm{CATALOG}],
\]

where CLS is the classification token representing the object itself, NORM is the normalization token that encodes the image standardization scale $s_i$, IMAGE contains tokens for NIRCam images, and CATALOG contains the tokens for the catalog data. The operator $\oplus$ means concatenation. Before training, the CLS token is initialized as an array of zeros, and the NORM token is computed using $s_i$ through a Multilayer Perceptron (MLP). The IMAGE and CATALOG tokens are constructed as follows:


\begin{enumerate}
    \item \emph{Image tokens.} Using similar ideas introduced in \citep{2024Smith}, for each combination of object + filters, we use $8\times8=64$ tokens to describe the image, where each token represents one $8\times8$ pixel patch. We choose this patch size to be larger than a typical PSF (about 4 pixels), but small enough to capture small-scale structures of large extended objects.  We use a simple MLP encoder to convert the input image patches into tokens. 
    This setup leads to 1,024 image tokens per object. 
    \item {\em Catalog tokens.} We first organize the catalog columns into 430 groups, following the rule listed in Table \ref{tab:catalog_tokens}. Briefly speaking, columns related to the same measurement (e.g., one object's flux and uncertainty in one filter and aperture) are put into the same group. Each group then enters an MLP encoder that returns the corresponding token.

\end{enumerate}


The total number of tokens describing an object and its associated data is thus 1,456. All tokens, regardless of what they represent, have the same length of 256 and are thus suitable for a transformer model (Section \ref{sec:architecture}).

\subsection{Transformer-based Architecture} \label{sec:architecture}

The main body of the model is a transformer, which consists of a series of Transformer blocks  \citep{Vaswai2017}. Each Transformer block performs multi-head cross-attention mechanisms to learn connections between different tokens. The input and output of the transformer have the same format, i.e., both consist of 1,456 tokens with lengths of 256. Throughout the transformer, the cross-attention mechanism learns the connection between different image patches, between different catalog columns, and between images and catalog data. The CLS token of each object receives information from all the other tokens, and becomes a representation of the object itself.

In addition to the main transformer, we add three MLPs with the same structure, all of which take the CLS token as input and return a 128-length vector. The three output vectors are named ${\bf z}_{\rm sci}$, ${\bf z}_{\rm img}$, and ${\bf z}_{\rm cat}$, respectively, which are used in the loss functions described below. We also include a lightweight image decoder to convert image tokens back to image patches, which is used in the image reconstruction loss.

\subsection{Loss Function} \label{sec:losses}

We consider three aspects when designing the loss function. First, we expect the model to reconstruct masked tokens from unmasked ones, following the idea of AION-1 \citep{Parker2025}. Second, the transformer should output similar embedding vectors for objects that look similar. Third, a regularization term should prevent the transformer from collapsing all objects onto a very low-dimensional space. To write a specific expression of the loss function, we first introduce several functions used in the loss. The smooth $L_1$ function is defined as \citep[e.g.,][]{fastrcnn}


%
%

\[
{\rm smooth}_{L1}(r)=
\begin{cases}
\tfrac12r^2, & |r|<1,\\
|r|-\tfrac12, & |r|\geq1
\end{cases}
\]

For $P,Q\in\mathbb{R}^{B\times d}$, we define the mean row-wise softmax cross-entropy function as

\begin{equation}
    \operatorname{CE}(P,Q)=\frac{1}{B}\sum_i^B Q_i \times \log \operatorname{softmax}(P_i)
\end{equation}

We then define the symmetric Information Noise-Contrastive Estimation (InfoNCE) loss for two 
 batches $U,V\in\mathbb{R}^{B\times d}$ as \citep{Oord2018}
\begin{equation}
\mathcal{I}_{\tau}(U,V)=\tfrac12\left[
\operatorname{CE}(UV^\mathsf{T}/\tau,I_B)+
\operatorname{CE}(VU^\mathsf{T}/\tau,I_B)\right],
\label{eq:infonce}
\end{equation}
where $I_B$ is the unit matrix with dimension $B$. For two vector batches $U$ and $V$, the InfoNCE loss decreases when $U$ and $V$ are better aligned. We use $\tau=0.07$ in this work.

We can now describe the three terms in the loss function:

\subsubsection{Reconstruction loss}
The reconstruction loss describes how well the model predicts masked tokens from unmasked tokens.
This loss forces the model to learn relations between different tokens (e.g., between images and catalogs, or between images of different bands).
For each object at each epoch, we randomly mask the tokens 
with probability 0.60 for image tokens and 0.25 for catalog tokens.
We then feed the unmasked tokens to the transformer and use MLP decoders to convert the transformer output to predict the masked image patches and catalog values. We then compute the reconstruction loss as
\begin{align}
\mathcal{L}_{\rm recon,token}
&=\mathcal{L}_{\rm SCI}+0.2\mathcal{L}_{\rm ERR}+\mathcal{L}_{\rm cat},\nonumber\\
{\rm where~} \mathcal{L}_{x}
&=\left\langle{\rm smooth}_{L1}(\widehat{x}-x)\right\rangle
\label{eq:image_loss}
\end{align}
where $\widehat{x}$ represents the reconstructed image or uncertainty patch from the image decoder, or the reconstructed catalog column values from the catalog decoder.

We also define the catalog reconstruction loss, where we mask all catalog tokens and keep all image tokens, then predict the masked catalog values using the catalog decoder:
\begin{equation}
\mathcal{L}_{\rm recon, cat}=
\left\langle{\rm smooth}_{L1}(\widehat{x}-x)\right\rangle
\label{eq:xmod_loss}
\end{equation}
where $\widehat{x}$ are the predicted catalog values of the image tokens. The total reconstruction loss is thus

\begin{equation}
    \mathcal{L}_{\rm recon}=\mathcal{L}_{\rm recon, token}+\mathcal{L}_{\rm recon, cat}
\end{equation}

\subsubsection{Alignment loss}

Alignment loss describes how well different views of an object align in the embedded space. Specifically, for the same object, two subsets of unmasked tokens should return similar embedded vectors, since the embedded vector represents the object itself. Similarly, the embedded vectors of two similar objects should also be similar. 

The detailed definition of alignment loss is as follows. For each object at each epoch, we first mask all catalog tokens to obtain the image-only view vector ${\bf z}_{\rm img}$, then mask all image tokens to obtain the catalog view vector ${\bf z}_{\rm cat}$ (Section \ref{sec:architecture}). We then define an image-catalog alignment loss:

\begin{equation}
    \mathcal{L}_{\rm align,img-cat}=
\mathcal{I}_{\tau}({\bf z}_{\rm img},{\bf z}_{\rm cat}).
\end{equation}

We also define a same-object alignment loss term. For each object at each epoch, we produce two different ``views'' (denoted as view A and view B) by randomly masking some tokens. We then take the science embeddings $({\bf z}_{\rm sci})$ of the two views, then define
\begin{equation}
\mathcal{L}_{\rm align, same-object}=\mathcal{I}_{\tau}({\bf z}_{\rm sci,A},{\bf z}_{\rm sci,B})
\end{equation}

Lastly, we define the cross-object term. The core idea is that two similar objects should have similar embeddings. To this end, we first define the ``distance" between objects $i$ and $j$ in the color and size space:
\begin{align}
d_{ij}^{2} = \frac{1}{K}\sum_{k=1}^{K}(x_{ik}-x_{jk})^2
\label{eq:science_dist}
\end{align}

where $x_{ik}$ represents colors and sizes of object $i$. In this work, we use five colors and one size as $x_{ik}$, including the 
 F115W--F444W, F150W--F444W, F200W--F444W, F277W--F444W, and F356W--F444W colors using the CIRC1 aperture magnitudes, as well as the $\log_{10}\mathrm{FWHM}$ where FWHM is the size of the object. These six columns are standardized using the median values and the interquartile range (i.e., the difference between the 75th and the 25th percentile) of the entire training sample. We then compute the similarity $s_{ij}$ that describes the similarity between object $i$ and $j$:

\begin{align}
s_{ij}&=\exp(-d_{ij}^{2}/2),\nonumber\\
\bar s_{ij}&=s_{ij}/\sum_l^B s_{il}
\label{eq:science_weights}
\end{align}

where the $l$ runs over objects within a batch.
Meanwhile, the similarity of the two objects can be computed using their science embeddings as $s'_{ij}=({\bf z}_{i,{\rm sci,A}})^{\mathsf{T}}{\bf z}_{j,{\rm sci,B}}$. For object $i$, we define the cross object alignment as the cross-entropy between $s_{ij}$ and ${s'_{ij}}$:
\begin{equation}
\mathcal{L}_{\rm align,cross-object}=\tfrac12\left[
\operatorname{CE}(s'_{ij}/\tau,\bar s_{ij})+
\operatorname{CE}(s'_{ji}/\tau,\bar s_{ij})\right]
\label{eq:metric_loss}
\end{equation}

The alignment loss is thus

\begin{align}
    \mathcal{L}_{\rm align} = &0.03\,\mathcal{L}_{\rm align,\,img\text{-}cat}\nonumber\\
    &+0.03\,\mathcal{L}_{\rm align,\,same\text{-}object}\nonumber\\
    &+0.05\,\mathcal{L}_{\rm align,\,cross\text{-}object}
\end{align}

\subsubsection{Regularization loss}

To prevent the transformer from dimensionality collapse \citep{jing21}, we adopt the Variance-Invariance-Covariance Regularization (VICReg) loss \citep{Bardes2022}. For objects in a batch of $B$, we denote their embeddings as $H\in\mathbb{R}^{B\times256}$, then define $\widetilde H=H-\langle H\rangle_B$ and $C=\widetilde H^\mathsf{T}\widetilde H/(B-1)$. We use superscript $v$ to denote different views (i.e., unmasked subsets of tokens) of one object. The VICReg loss is

\begin{align}
\mathcal{L}_{\rm var}
&=\frac{1}{2}\sum_{v=1,2}\frac{1}{256}\sum_j
\max\!\left[0,\,1-\sqrt{\operatorname{Var}\big(H^{(v)}_{:j}\big)+10^{-4}}\right],\nonumber\\
\mathcal{L}_{\rm cov}
&=\frac{1}{2}\sum_{v=1,2}\frac{1}{256}\sum_{j\ne k}\big(C^{(v)}_{jk}\big)^{2},\nonumber\\
\mathcal{L}_{\rm inv}
&=\frac{1}{256\,B}\sum_{i=1}^{B}\sum_{j=1}^{256}
\big(H^{(1)}_{ij}-H^{(2)}_{ij}\big)^{2}\nonumber\\
\mathcal{L}_{\rm VICReg}&=0.02\mathcal{L}_{\rm var}+0.02\mathcal{L}_{\rm inv}+0.001\mathcal{L}_{\rm cov}
\label{eq:vicreg_loss}
\end{align}

The total loss is thus
\begin{align}
\mathcal{L}=\mathcal{L}_{\rm recon}+\mathcal{L}_{\rm align}+\mathcal{L}_{\rm VICReg}
\label{eq:total_loss}
\end{align}

\subsection{Training Results} \label{sec:training}

We randomly initialize the model and train from scratch for 20 epochs. 
We use AdamW optimizer \citep{Loshchilov2019} with a constant learning rate of $3\times10^{-4}$. 
The batch size is $B=256$. 
The training loss decreases steadily before epoch 10 and then gradually flattens. The reconstruction loss contributes about 80\% of the total loss.

\begin{figure*}
\centering
\includegraphics[width=1\linewidth]{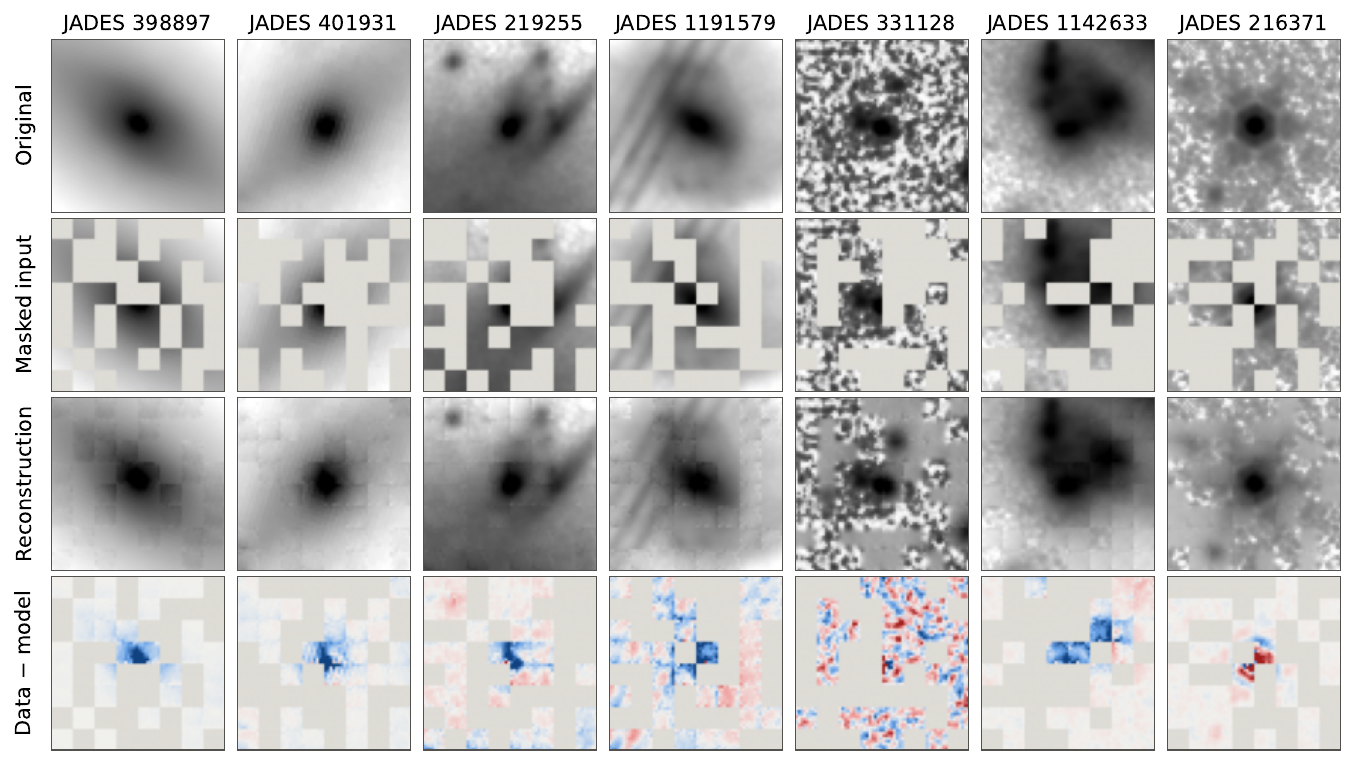}
\caption{Masked image patch reconstruction for seven illustrating examples. From top to bottom: the original image, the masked image, the reconstructed image, and the reconstruction residual (red means positive). For each object, we mask half of the F277W patches, while the other F277W patches, all other available bands, and the catalog remain visible. This illustrates that the transformer learns the relation between the masked image patches and successfully reconstructs the overall morphology.}
\label{fig:recon}
\end{figure*}

To demonstrate that the model successfully learns the connections between tokens,
in Figure \ref{fig:recon}, we show several examples of image token reconstruction. For these objects, we mask half of the image tokens in the F277W band, while keeping the other tokens (image and catalog) unmasked. The reconstruction restores the main structures of the masked patches. 
We note that the model is not trained as a dedicated image reconstruction tool, and Figure \ref{fig:recon} only demonstrates that the model extracts information about the connections between tokens. One noticeable caveat is that the model underestimates bright pixels, likely due to the saturated response of the asinh function at high fluxes (Equation \ref{eq:image_transform}). A dedicated image reconstruction tool will need specific fine-tuning (e.g., for model architecture and pixel weights) and is out of the scope of this study.

The model also learns to extract morphological information from images that is not present in the catalog. We demonstrate this by evaluating the correlation between the summary embedding and two non-parametric morphological indicators, namely the concentration $(C)$ and the asymmetry $(A)$. We measure these quantities directly from the F444W image (or from the reddest available band when F444W is unavailable):
\begin{align}
C_{6/16}
&=\frac{\sum_{r_p\leq6}|I_p|}{\sum_{r_p\leq16}|I_p|},\nonumber\\
A_{180}
&=\frac{\sum_p\left||I_p|-|I_{R_{180}(p)}|\right|}
{2\sum_p|I_p|}.
\label{eq:morphology_proxies}
\end{align}
where $I_p$ denotes the image, $r_p$ denotes the distance from pixel $p$ to the cutout center in units of pixels, and $R_{180}$ is the operation of rotating the image by 180$^\circ$. To avoid leakage from catalog information, when constructing the embedding vectors, we only use image tokens as input and mask all catalog tokens.

\begin{figure*}
\centering
\includegraphics[width=0.8\linewidth]{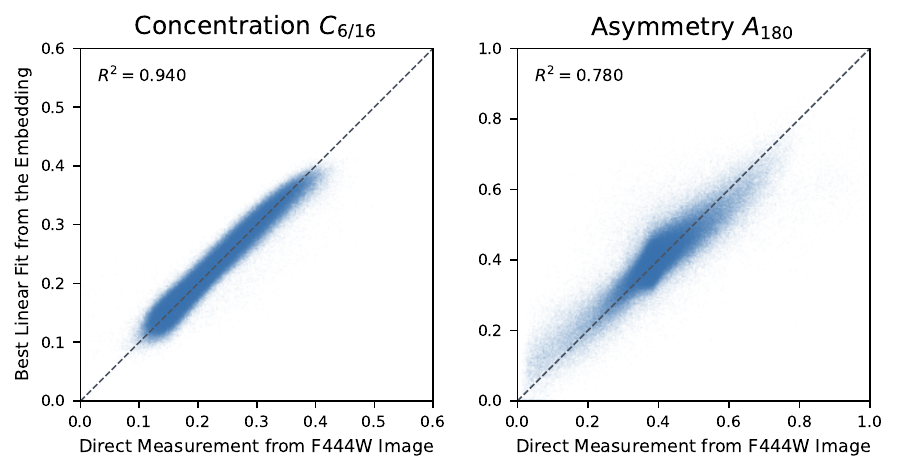}
\caption{Linear regression between the CLS embedding vectors of objects and the non-parametric morphology measurements, concentration $(C_{6/16})$ and asymmetry $(A_{180})$. We measure $C_{6/16}$ and $A_{180}$ using the F444W band image, then perform a linear regression between the CLS vector and $C_{6/16}$ and $A_{180}$. The $x-$axis is the direct measurement from F444W images, and the $y-$axis is the best-fit value from the linear regression. The CLS vector strongly correlates with $C_{6/16}$ and $A_{180}$, indicating that the model learns morphological information of objects, even if we do not include such information in the input catalog. }
\label{fig:morphology}
\end{figure*}

Figure \ref{fig:morphology} shows the correlation between the CLS vector and the two quantities ($C_{6/16}$ and $A_{180}$). We fit $C_{6/16}$ and $A_{180}$ as linear functions of the CLS vector and compare them with the values measured from the F444W image. We also compute the correlation coefficient (R$^{2}$ = 0.937 for $C_{6/16}$, R$^{2}$ = 0.768 for $A_{180}$) between the values predicted by CLS and the values measured by the image. 
Figure \ref{fig:morphology} and the correlation coefficient indicate a strong correlation between the $C_{6/16}$ and $A_{180}$ values predicted from the CLS vectors and measured by the image. This suggests that the CLS vectors encode their morphology information directly from images, without information from the catalog tokens.


\section{Blind Active Discoveries of Rare Astrophysical Populations}
\label{sec:blind_discovery}

One of the main goals of FM-JADES-v1 is \emph{blind active discovery}:
instead of waiting for unusual sources to be recognized serendipitously during visual inspection or searching with predefined population-specific criteria, we systematically search the learned representation for detached and internally coherent structures whose astrophysical identities are not specified in advance.

\subsection{A Label-independent Search for Embedding Islands}

To identify detached subsamples in the JADES sample, we project the learned CLS embedding of objects onto a two-dimensional space using UMAP \citep{umap}.
To reduce the effect of non-uniform coverage, we use a common, deliberately controlled subsample by requiring coverage in all seven photometric broad bands of NIRCam (F090W, F115W, F150W, F200W, F277W, F356W and F444W) and CIRC1 S/N$>10$ in both F356W and F444W, yielding 122,336 sources. We refer to this sample as the ``UMAP sample" in the following text. The model receives only these seven images and the corresponding catalog information, including multi-aperture photometry, half-light radii, and colors. 
The resulting 256-dimensional CLS embedding therefore represents a uniform seven-band view of the survey with jointly encoding imaging and photometric information.
The embeddings are $L_2$-normalized and projected with cosine UMAP using 100 neighbors and a minimum-distance parameter of 0.10. 

\begin{figure}
\centering
\includegraphics[width=1\columnwidth]{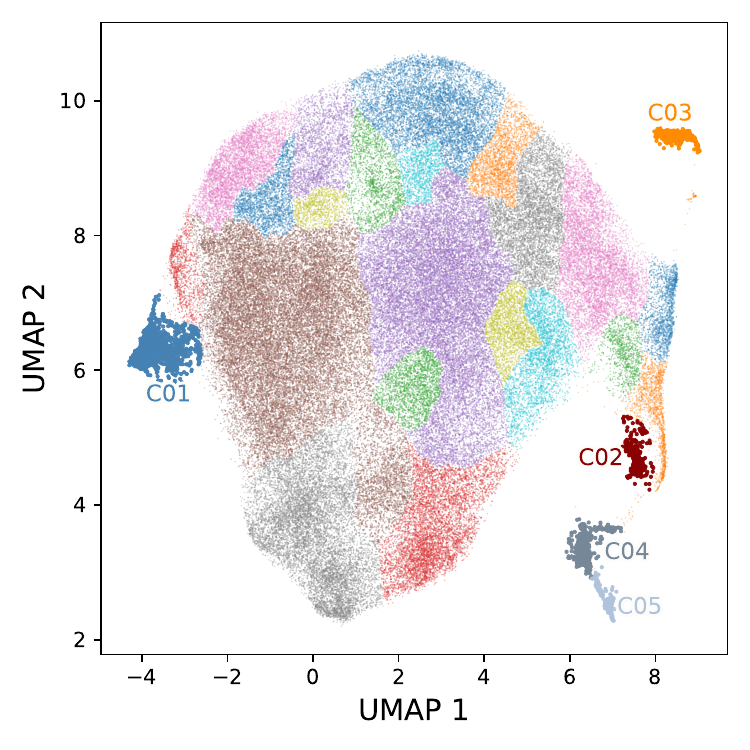}
\caption{The UMAP projection of 122,336 objects with uniform band coverage and high SNR (see section \ref{sec:blind_discovery} for details). The UMAP consists of a mainland (shown in small points, with different colors marking different HDBSCAN clusters) and five detached islands (shown in large points and marked by text). C01 (high-redshift galaxies), C02 (red compact objects like LRDs), C03 (bright point sources) have distinct observational features, while C04 and C05 are dominated by artifacts (e.g., image edges and PSF spikes).}
\label{fig:umap}
\end{figure}

Figure \ref{fig:umap} shows the projected UMAP.
We run HDBSCAN for the 2-dimensional UMAP vectors using the \texttt{scikit-learn} \citep{scikit-learn} package to identify clusters. Points initially classified as transition or noise are assigned to nearby clusters using a distance-weighted five-nearest-neighbor rule. 
This analysis returns 29 density clusters, and visual inspection (VI) suggests that 24 clusters form a mainland, and five clusters are detached from the mainland. We name the five clusters C01 to C05, as marked in Figure \ref{fig:umap}. 
The HDBSCAN clustering result yields a low mean Silhouette score of 0.120 for the ``mainland" clusters, while the five islands have substantially higher Silhouette scores (0.6 to 0.9). The discussion above indicates that the five islands are well separated from the mainland both visually and quantitatively.  

To further demonstrate that the five clusters are statistically isolated, we evaluate the saddle-to-peak ratio for HDBSCAN-identified clusters. In short, the saddle-to-peak ratio quantifies how strongly a cluster is connected to its neighbors. 
The specific analysis is as follows.
We first evaluate the object number density on the UMAP by building a 2D histogram with a pixel size of 0.03, then smoothing the histogram using a Gaussian kernel with $\sigma=1$ pixels. We then measure the saddle-to-peak density ratio $\lambda$ for each cluster using \texttt{astrodendro} \citep{Rosolowsky08}. 
 The five islands have $\lambda<0.06$ while all other clusters have ratios larger than $\lambda>0.25$, meaning that the five islands are statistically more isolated than other clusters.

The five detached clusters represent anomaly populations in JADES that have significantly different features compared to the majority of objects.
We emphasize that the clustering analysis is unsupervised, which leads to the central idea of the {\em blind active discovery} framework: we first identify structures in the embedding space without any pre-defined selection, then analyze the astrophysical properties of the identified structures. This process is ideal for the discovery of previously unknown populations. In the rest of this Section, we will discuss the properties of each island and which type of objects they correspond to.




\subsection{Characterizing the Blindly Discovered Islands}

For each island, we visually inspect the images and SEDs of a random subset of 50 sources. We also match the dataset to the DAWN {\em JWST} Archive (DJA)\footnote{https://cosmicdawn.dk/dja/} to identify objects with NIRSpec spectroscopy and inspect them. We find that C04 (465 objects) and C05 (477 objects) are dominated by artifacts and bad image quality. C01, C02, and C03 represent observationally distinct populations:

\begin{enumerate}
    \item C01 (1,249 objects) is dominated by high photometric redshift $(z\gtrsim7)$ objects, with many spectroscopically-confirmed high-redshift galaxies;
    \item C02 (347 objects) contains red, compact objects, with many of them being published LRDs;
    \item C03 (569 objects) is a point source island dominated by galactic stars. Among the 17 objects in C03 with NIRSpec/Prism spectra, 16 are classified as stars. 
\end{enumerate}

This work focuses on the application of FM-JADES-v1 in extragalactic studies. Below we investigate C01 and C02 islands in detail.



\subsection{C01: The High-redshift Island}
\label{sec:C21}
\begin{figure*}
    \centering
    \includegraphics[height=0.4\linewidth]
    {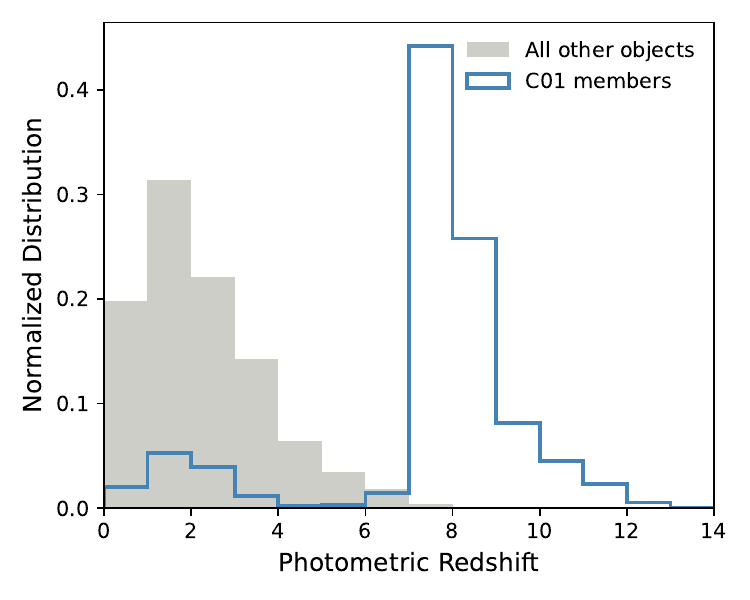}
    \includegraphics[height=0.45\linewidth]
    {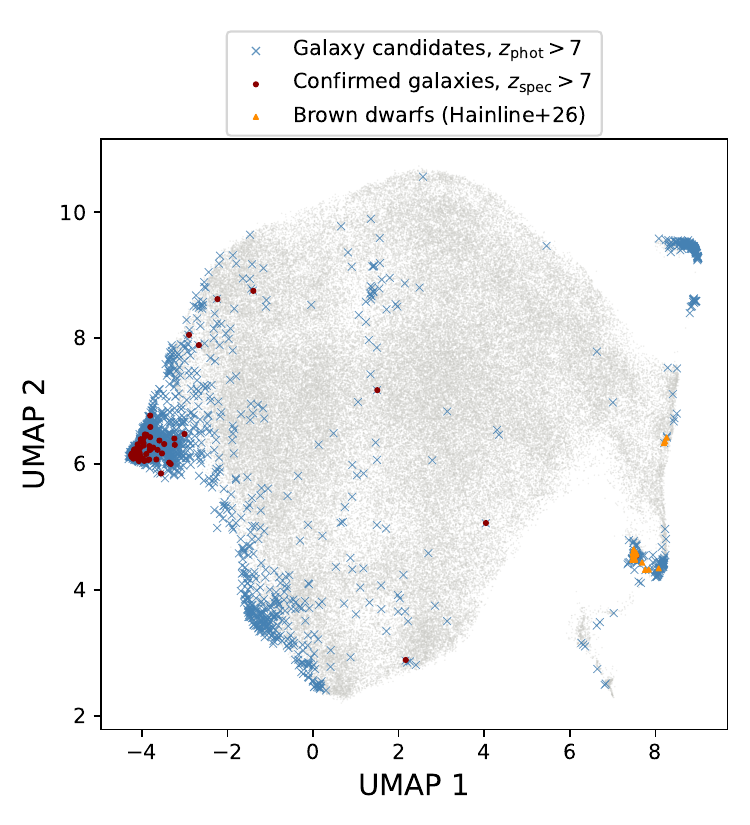}
    \caption{Left Panel: The normalized photo-$z$ distribution of C01 island. C01 is dominated by objects with photo-$z$ $>$ 6. Right Panel: The positions of high redshift galaxies on the UMAP. Blue crosses mark galaxies with $z_{\rm phot}>7$, the red dots mark spectroscopically confirmed $z_{\rm spec}>7$ galaxies, and the orange triangles mark brown dwarfs in \cite{Hainline2026}. C01 contains 1,068 out of 1,891 $z_{\rm phot}>7$ galaxies and 91 out of 100 $z_{\rm spec}>7$ galaxies. The brown dwarfs are clearly separated from spectroscopically confirmed high redshift galaxies on this plot.}
    \label{fig:c01photoz}
\end{figure*}

C01 contains 1,249 members with a median $z_{\rm phot}=7.52$, $m_{\rm F444W}=28.3$, FWHM $=0\farcs187$, and F200W$-$F444W$=0.48$. 
The left panel of Figure \ref{fig:c01photoz} shows the normalized distribution of the photometric redshift for C01 objects and for the rest of the sample. Objects in C01 exhibit systematically higher photo-z distributions than those in other groups, with 517 (1,068) of the 1,249 objects in C01 having $z_{\rm phot}>8$ ($z_{\rm phot}>7$). 
We further obtain the JADES DR4 spectroscopic galaxy sample \citep{CurtisLake2026,Scholtz2026}, and take grade A and B as reliable spectroscopic redshifts. The seven-band high SNR sample contains 100 $z_{\rm spec}>7$ galaxies, 91 of which are located in the C01 island. We therefore conclude that C01 traces the high redshift galaxy population. We note that C01 has a tail of low redshift objects with median $z_{\rm phot}\sim2$, which are likely low redshift quiescent galaxies with a Balmer break mimicking the Ly$\alpha$ break of high-redshift galaxies.

\begin{deluxetable*}{lllccc}
\tablecaption{Benchmark of blind rare-population discovery
\label{tab:discovery_benchmark}}
\tablehead{
\colhead{Population} &
\colhead{Reference sample} &
\colhead{Island} &
\colhead{$N_{\rm island}/N_{\rm total}$} &
\colhead{Recovery} &
\colhead{Enrichment}
}
\startdata
$z>7$ galaxy & Photometric    & C01 & 1068/1891   & 56.5\%   & 55$\times$ \\
$z>7$ galaxy & Spectroscopic & C01 & 91/100   & 91.0\%   & 89$\times$ \\
$z>8$ galaxy & Photometric    & C01 & 517/912   & 56.7\%   & 55$\times$ \\
$z>8$ galaxy & Spectroscopic & C01 & 33/37   & 89.2\%   & 87$\times$ \\
LRD          & Photometric    & C02 & 109/181 & 60.2\% & 212$\times$ \\
LRD          & Spectroscopic & C02 & 25/35   & 71.4\% & 252$\times$ \\
\enddata

\tablecomments{
Population labels are introduced only after the embedding and island
definitions are finalized. Recovery is defined as the fraction of objects in
each reference sample inside the corresponding island. Enrichment is the population fraction within the island divided by
its fraction in the full 122,336-source sample.
}
\end{deluxetable*}

\begin{figure*}
\includegraphics[width=\textwidth]{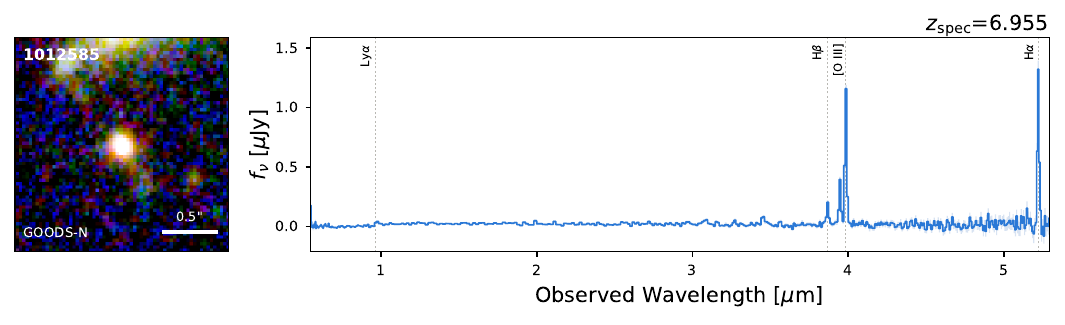}
\caption{A previously-unreported galaxy at $z=6.955$ with NIRSpec/Prism observation on C01.}
\label{fig:stamp_highz}
\end{figure*}

The right panel of Figure \ref{fig:c01photoz} shows the distribution of objects with $z_{\rm phot}>7$ and $z_{\rm spec}>7$ on the UMAP. The high concentration ($91/100=0.91$) of $z_{\rm spec}>7$ galaxies in C01, compared to that of the $z_{\rm phot}>7$ objects ($1068/1891=0.56$), suggests that the embedding encodes information about how likely an object is a real high-redshift galaxy, in addition to the photometric redshift. The details of the recovery and enrichment of the high-redshift galaxies in C01 are listed in Table \ref{tab:discovery_benchmark}. Also, we mark the positions of brown dwarfs from \cite{Hainline2026} on the UMAP, which are a common population of contamination for high-redshift galaxy searches. The right panel of Figure \ref{fig:c01photoz} shows that the UMAP clearly separates brown dwarfs from the C01 island. Therefore, the embedding might be used to increase the purity of the high redshift galaxies candidates. In C01, we also find one unpublished galaxy at $z=6.955$ with DJA spectra (Figure \ref{fig:stamp_highz}). 

\begin{figure}
\centering
\includegraphics[width=0.9\linewidth]{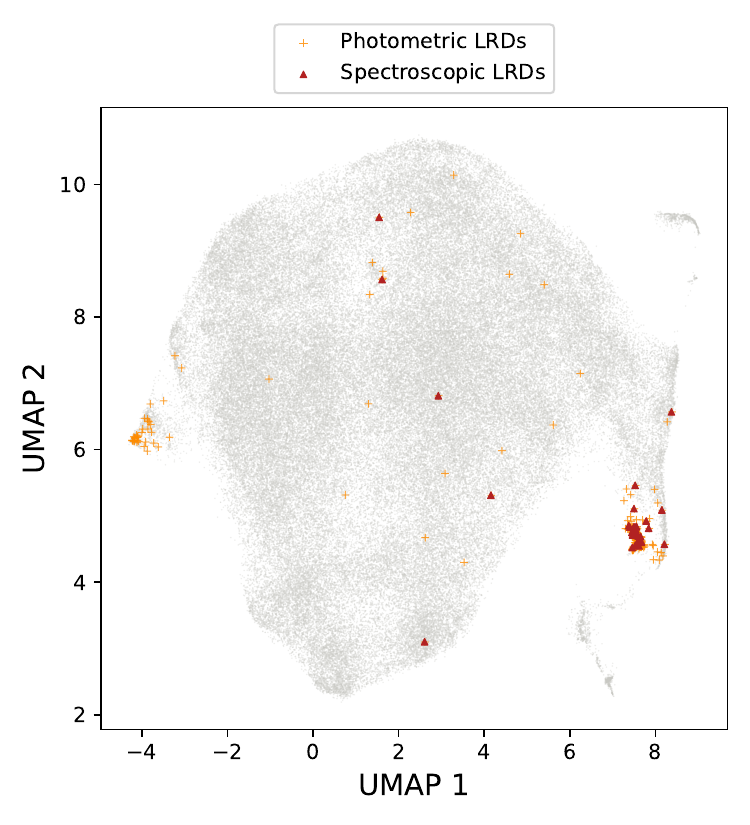}
\caption{The distribution of LRDs on the UMAP. Orange crosses mark photometrically selected LRDs and red triangles mark spectroscopically confirmed ones. C02 contains 109 out of 181 of the photometrically selected LRDs and 25 out of 35 spectroscopically confirmed LRDs.}
\label{fig:umap_highz_lrd}
\end{figure}

\subsection{C02: The LRD-rich Island}
\label{sec:C22}
C02 contains 347 objects with a median $z_{\rm phot}=5.28$, $m_{\rm F444W}=26.2$, FWHM $=0\farcs170$, and F200W$-$F444W$=1.96$. C02 is highly enriched by LRDs. To demonstrate this, we cross-match the 122,336 objects in the UMAP sample with literature, and identify 181 LRDs in the sample, among which 35 are spectroscopically-confirmed\footnote{We define an LRD to be ``spectroscopically-confirmed" only if the reporting paper claims the object as an LRD based on its spectrum. Some photometrically-selected LRDs have spectra, but their spectra indicate that these objects are indeed not LRDs. We do not count such objects as spectroscopically-confirmed LRDs.} \citep{Matthee2024,Maiolino2023,Williams2023,PerezGonzale2024,Kokorev2024,Kocevski2024,Feeney2024,Rinaldi2025,Barro2026,Lin2026,Zhang2025NarrowLine,deGraaff2025,Barro2026AGN,Liu2026,Weibel2026}. C02 recovers 109 LRDs in total and 25 spectroscopically confirmed ones. These benchmark results are listed in Table \ref{tab:discovery_benchmark}.

The C02 clusters contain 238 objects that have not been reported as LRDs, and 23 of them have been reported as brown dwarfs by \cite{Hainline2026}. We visually inspect the images and SEDs of the remaining 215 objects, finding that most of them are compact sources with red colors. These objects might be previously unreported LRDs (see Section \ref{sec:blind_discovery}), strong emission line galaxies where the emission lines produce the red color, or compact quiescent galaxies with a strong Balmer break. 
A detailed classification for these objects requires follow-up spectroscopy and is out of the scope of this study.

Figure \ref{fig:umap_highz_lrd} shows the distribution of photometric and spectroscopic LRDs on the UMAP. Interestingly, a population of photometrically-selected LRDs is concentrated in C01, but none of them are spectroscopically confirmed as LRDs. We investigate the 11 LRDs on C01 that have JADES DR4 spectra, finding that they are indeed high-redshift emission line galaxies. This result indicates that photometrically selected LRDs form a composite sample of many different populations, consistent with findings of recent studies \citep[e.g.,][]{Rinaldi2025,hainline25b,pg26,Ginolfi2026}.

\subsection{From Blind Discovery to Candidate Retrieval}

The natural next experiment is to investigate whether the embeddings can efficiently retrieve additional objects once an astrophysical population has been identified. 
We perform this experiment by selecting 19 LRDs in the UMAP sample with unambiguous broad line detections. Using these LRDs, we perform a reverse-$k$th Nearest Neighbor \citep[rkNN,][]{rKNN} search through the UMAP sample. Specifically, for each object in the UMAP sample, we retrieve the 20 nearest nonself neighbors (adopting cosine distance) using their 256-dimension CLS embedding. If there is at least one broad-line LRD neighbor, we classify the object as a candidate. In addition, we require the candidate to be located on the LRD-rich island C02.

We identified 113 candidates for LRDs from the 19 seed LRDs. Among the 113 candidates, 57 were previously unreported. This selection recovers $56/109=51.4\%$ of the photometrically-selected LRDs in C02, and $20/25=80\%$ of the spectroscopically-selected LRDs.  This experiment shows the feasibility of recovering a large sample of objects from a small number of seeds by simple similarity search. This is crucial for new population studies when there are only few objects available; similarity search allows us to expand the sample efficiently without fine-tuning specific selection criteria like color cuts. We also note that the selection is not expected to be complete, as the 19 seed LRDs might not be sufficiently representative enough for the entire LRD population.

\begin{figure*}
\includegraphics[width=\textwidth]{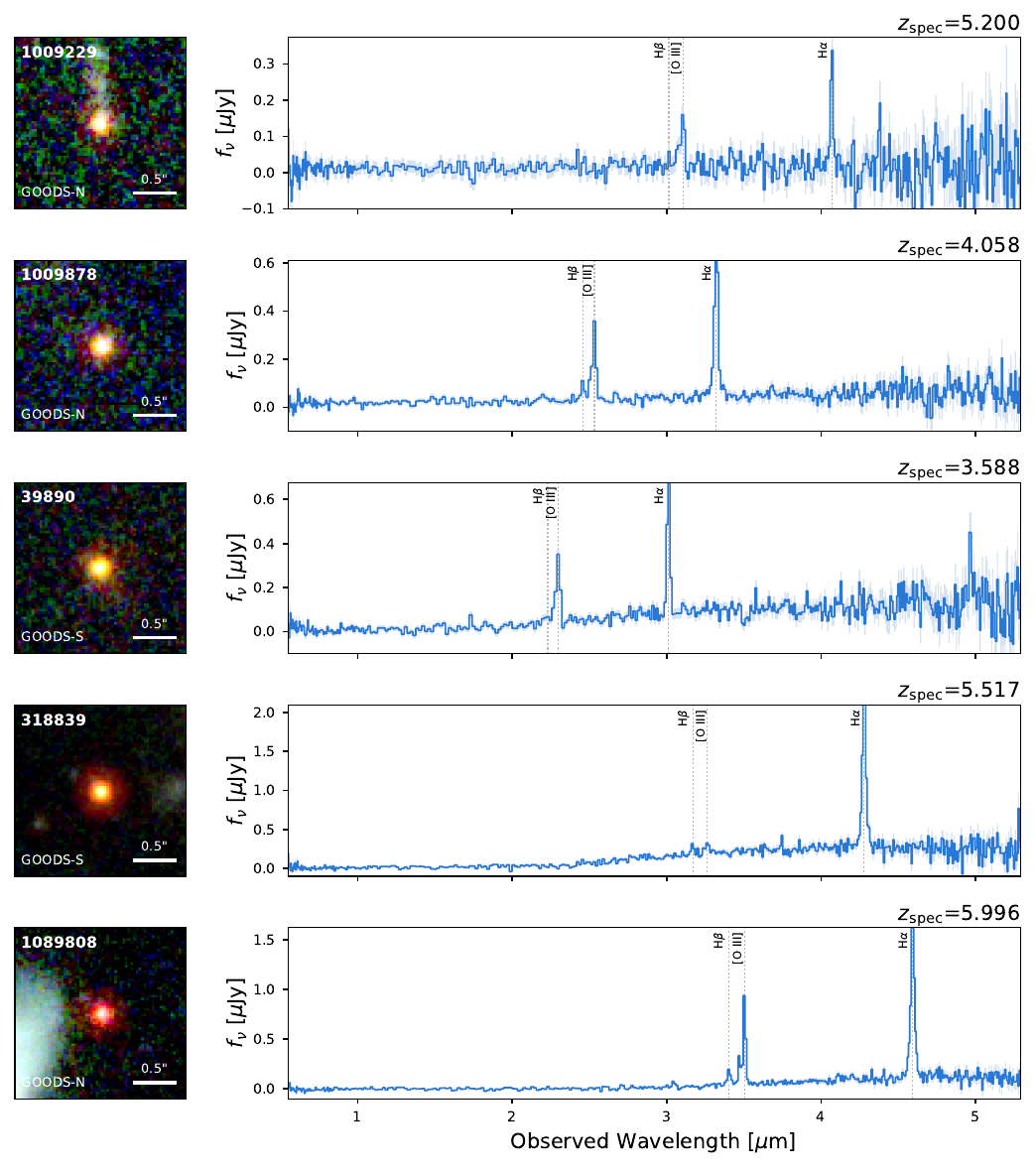}
\caption{Top 5 rkNN-ranked previously-unreported LRDs with NIRSpec/PRISM observations. We use 19 LRDs with spectroscopically-confirmed broad lines as seeds for the rkNN. We only include objects with full NIRSpec/Prism wavelength coverage. We note that ID 39890 and ID 318839 were reported by \cite{Barrufet25} but are not classified as LRDs. The full list of newly-identified LRD candidates can be found in Table \ref{tab:cands}.}
\label{fig:stamps}
\end{figure*}

Finally, we find that eight of the 57 unreported candidates have DJA NIRSpec/Prism spectra. Figure \ref{fig:stamps} shows five of them; we exclude the other three objects due to significant incomplete spectral coverage. All five objects show H$\alpha$ emission lines and LRD-like red continua. Follow-up high-resolution NIRSpec observations can determine whether they exhibit broad H$\alpha$ components. The full candidate list can be found in Table \ref{tab:cands}.

\begin{deluxetable*}{lcccccc}
\tablecaption{LRD candidates in the C02 island using rkNN selection \label{tab:cands}}
\tablehead{
\colhead{ID} & \colhead{R.A.} & \colhead{Decl.} &
\colhead{$m_{\rm F200W}$\tablenotemark{a}} & \colhead{$m_{\rm F356W}$\tablenotemark{a}} &
\colhead{$m_{\rm F444W}$\tablenotemark{a}} & \colhead{Photometric Redshift} \\
\colhead{} & \colhead{(deg)} & \colhead{(deg)} &
\colhead{(mag)} & \colhead{(mag)} & \colhead{(mag)} & \colhead{}
}
\startdata
1006892 & 189.12434 & +62.24589 & 27.21 & 25.98 & 26.12 & 4.40 \\
216374  &  53.13259 & --27.75240 & 25.95 & 24.18 & 24.41 & 4.34 \\
1009878\tablenotemark{*} & 189.21153 & +62.25534 & 27.85 & 26.38 & 26.54 & 4.32 \\
1213348 & 189.40352 & +62.26207 & 26.09 & 24.31 & 24.27 & 4.34 \\
1214070 & 189.42584 & +62.28927 & 26.50 & 24.28 & 24.34 & 4.33 \\
1135412 & 189.43025 & +62.27379 & 28.29 & 26.13 & 25.46 & 5.01 \\
1136964 & 189.37383 & +62.28085 & 27.97 & 27.16 & 25.91 & 5.21 \\
1009229\tablenotemark{*} & 189.20622 & +62.25295 & 28.77 & 27.70 & 26.53 & 5.18 \\
\enddata
\tablenotetext{a}{The CIRC0 magnitude in JADES DR5 Catalog.}
\tablenotetext{*}{Have NIRSpec/Prism observation in DJA.}
\tablecomments{The full table will be available online after paper accepted.}
\end{deluxetable*}

\section{Improvement for Fewer-band photometric redshifts} \label{sec:photozapp}

As a foundation model, FM-JADES-v1 provides information-rich embeddings for objects and enables a wide range of downstream tasks. As an example, in this Section, we describe how to use FM-JADES-v1 to improve photometric redshift estimates when only a small number of bands are available. 
This task is particularly important for transferring information in deep, multi-band imaging surveys into other datasets. For example, NIRCam grism surveys, like Emission-line galaxies and Intergalactic Gas in the Epoch of Reionization \citep[EIGER;][]{Kashino2023EIGER}, A SPectroscopic survey of biased halos In the Reionization Era \citep[ASPIRE; ][]{Wang2023ASPIRE}, and Slitless Areal Pure-Parallel High-Redshift Emission Survey \citep[SAPPHIRES;][]{Sun2025}, often have images in only two or three bands. The Roman High-Latitude Wide-Area Survey (HLWAS) will only deliver F106, F129 and F158 images in the $\sim2,400$ deg$^2$ medium tier \citep{ROTAC2025}. Traditional photo$-z$ methods are based on template fitting and will fail due to severe degeneracies. In contrast, FM-JADES-v1 can extract objects' flux and morphology information that are correlated with redshift, significantly improving photo$-z$ estimates with few bands.

\begin{figure*}[!t]
\centering
\includegraphics[width=0.8\linewidth]{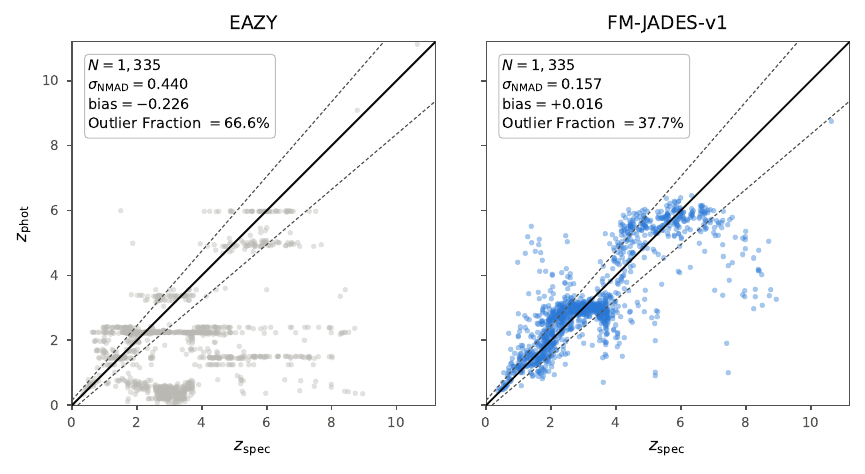}
\caption{Three-band photometric-redshift predictions versus
spectroscopic redshifts for the common 1,335 object test set.
Left Panel: The photo-$z$ measurement from EAZY template fitting. 
Right panel: The frozen pre-trained encoder followed by the post-trained attention-pooling head. Both photo-$z$ measurements use F115W,
F200W, and F356W. 
The solid line marks
$z_{\rm phot}=z_{\rm spec}$, and dashed lines mark
$|\delta z|=0.15$, where
$\delta z=(z_{\rm phot}-z_{\rm spec})/(1+z_{\rm spec})$. FM-JADES-v1 shows significant improvement in photo$-z$ measurements compared to traditional template fitting, as the embedded vectors encode flux and morphology information.
\label{fig:photoz}}
\end{figure*}

To demonstrate this, we obtain the JADES DR4 spectroscopic catalog \citep{Bunker2024,DEugenio2025,CurtisLake2026,Scholtz2026}, and select objects with quality A and B redshift measurements. This leads to a sample of 2,772 objects. We then obtain their Kron magnitudes in F115W, F200W, and F356W bands, matching the filters used in the EIGER and ASPIRE programs, and perform photo$-z$ measurements using EAZY \citep{Brammer2008}. The result is shown in the left panel of Figure \ref{fig:photoz}; we note that we restrict the sample to 1,335 objects that match FM-JADES-v1's test sample described below. The scatter of ${\rm d}z/(1+z)$ is $\sigma_{\rm NMAD}=0.44$ for the EAZY prediction. The outlier fraction, defined as the fraction of objects with $|z_{\rm phot}-z_{\rm spec}|>0.15\times(1+z_{\rm spec})$, is 66.6\%. The redshifts of many objects remain effectively unconstrained, as multiple templates at different redshifts give $\chi^2\approx0$.

We then test the performance of FM-JADES-v1 in the same task. This is done in two steps. In the first step, we attach a post-training attention-pooling head to the transformer. We fix the pre-trained model weights, use image and catalog tokens related to F115W, F200W, and F356W bands as input, and fit the head output to $\ln(1+z_a)$ from the catalog. 
We fit all objects in the JADES sample except for the 2,772 spectroscopic-redshift objects. The goal of this step is to distill redshift-related information from the three bands. In the second step, we re-fit the attention-pooling head using half of the spectroscopic-redshift sample, and evaluate the model using the other half of the spectroscopic-redshift sample. The right panel of Figure \ref{fig:photoz} shows the result. This model gives $\sigma_{\rm NMAD}=0.156$ for ${\rm d}z/(1+z)$, as well as an outlier fraction of 37.7\%, significantly better than the EAZY baseline.

Our result indicates that the fluxes and morphology of objects encode crucial information about galaxies' redshifts, and FM-JADES-v1 is able to extract such information. The three-band photo$-z$ will be particularly useful in determining the redshifts of single-line emitters (like H$\alpha$ and Pa$\alpha$ emitters) in NIRCam grism surveys, which have very similar features in their 2D grism spectra. Note that transferring this model to surveys with different depth may need specific fine tuning/post-training using the data from the different survey, which is out of the scope of this work.

\section{Discussion} \label{sec:discussion}

\subsection{Comparison with related work} \label{sec:related}

Self-supervised representation learning has shown strong and diverse applications in astronomical surveys, from contrastive learning on SDSS and DECaLS imaging \citep{Hayat2021,Stein2021} to image and spectrum alignment with AstroCLIP \citep{Parker2024}, as well as heterogeneous multi-modal modeling with AION-1 \citep{Parker2025}. \textsc{FM-JADES-v1} is designed for a more specific scientific workflow in deep {\em JWST} fields with a focus on using the learned representation as a discovery space for rare or previously unspecified populations. This model can also transfer the information learned from deep multi-band data to photo-$\it{z}$ measurement when only a few bands are available in other surveys.

An interesting finding in our work is that photometrically selected LRD show two clusters in the embedded space, one on the high-redshift island (C01) and one on the main LRD island (C02).
Recent manifold learning work has similar results showing that LRDs dominated two compact regions in the {\em JWST} photometric UMAP space, with part of the separation related to the redshift dependence of their observed colors \citep{Ginolfi2026}. Our analysis provides a label-independent test of this result. Also, our framework mainly uses the learned embedding vectors for data analysis, with UMAP serving only as a low-dimensional projection to reveal potentially interesting structures. The majority of downstream tasks, like similarity searches and photometric redshift improvements, are performed in the embedding space.


\subsection{Limitation}
Despite its potential discussed above, the current framework has several important limitations. First, \textsc{FM-JADES-v1} is trained on heterogeneous JADES data, and its learned representation can be sensitive to missing photometric bands and spatially varying depths. In such cases, the model may encode survey systematics rather than intrinsic astrophysical differences. Second, the learned representation is shaped by the training object distribution, so populations that are poorly sampled by only a few tens of examples in JADES may not exhibit reliable structures in the embedding. This limits our representative learning sensitivity to extremely rare populations.
Third, structures in low-dimensional projections such as UMAP should not be interpreted as distinct physical populations without further validation, since the projection can distort distances and emphasize local structures. Finally, the learned embeddings should be used to identify unusual populations, while additional physical measurements, spectroscopy, and follow-up analyses are still required to determine the astrophysical origin of the structures identified by the model. Addressing these limitations will require more heterogeneous training data, further refinement of missing modalities and selection effects, as well as systematic validation across independent surveys. 

These limitations also highlight where the current framework is most applicable. \textsc{FM-JADES-v1} is particularly useful for large and relatively homogeneous surveys such as Rubin LSST, Roman, and Euclid. These surveys with consistent observing strategies and uniformly reduced data products can reduce the risk that the model encodes survey artifacts as an intrinsic astrophysical structure. Such datasets provide ideal tests for scaling active discoveries to orders of magnitude larger dataset.

A key next step is therefore to extend the framework to heterogeneous datasets, where missing bands, non-uniform depth, variation in image quality, and survey-to-survey systematics must be explicitly modeled.

\section{Conclusion} \label{sec:summary}

We present \textsc{FM-JADES-v1}, a self-supervised multi-modal foundation model that learns a joint representation of {\em JWST} imaging and catalog measurements from JADES. The resulting embedding encodes both photometric and morphological information and can be reused for multiple scientific tasks without retraining the encoder. Our main results are as follows.

\begin{enumerate}

\item {The learned representation retains physically meaningful information beyond the explicit catalog input.} 
Probes of the image-only embedding recover morphological structure, including central concentration and asymmetry, demonstrating that the representation captures information from the resolved imaging that is complementary to broadband photometry.

\item {The representation enables blind active discoveries of rare populations.} 
We develop a label-independent search that systematically identifies detached and internally coherent structures in the embedding space, without relying on predefined population selections or serendipitous visual inspection.  
This search can be followed by inspecting the shared astrophysical properties of the population in the detached structure and characterizing them with available physical measurements. 
Applied to a uniform seven-band JADES sample of 122,336 sources, we identify two clusters corresponding to high-redshift galaxies (C01) and LRDs (C02), demonstrating the feasibility of blind active discovery for new populations.

\item {We benchmark blind active discovery of high-redshift galaxies and LRDs.}
We define new benchmark tests that evaluate how strongly the rare objects are clustered in the embedded space. Specifically,
we measure the recovery and enrichment of high-redshift galaxies and LRDs in C01 and C02 clusters, which are listed in Table \ref{tab:discovery_benchmark}. Both high-redshift galaxies and LRDs are highly enriched ($>50\times$ compared to the entire sample) in their corresponding islands. These numbers can be compared with those of other models that work on the JADES dataset. 

\item {The same representation enables efficient generalization from discovered populations to new candidates.} 
Using only 19 secure broad-line LRDs as seeds, a rkNN
search identifies 113 candidates in C02, recovering 51.4\% of the
photometrically selected and 80\% of the spectroscopically selected LRDs. It also identifies 57 candidates that have not been reported in the literature for follow-up. Public NIRSpec/Prism
spectra for five well-covered candidates show H$\alpha$ emission and LRD-like continuum (see Figure \ref{fig:stamps}), illustrating how similarity search can rapidly expand rare
populations. 

\item {The learned representation can transfer information to fewer-band photometric measurements.} 
In a controlled three-band experiment using only F115W, F200W, and F356W imaging and the corresponding catalog tokens, a lightweight attention output on the frozen encoder reaches $\sigma_{\rm NMAD}=0.157$, compared with 0.440 for three-band EAZY on the same test sample. This result demonstrates that representations learned from information-rich survey data can provide useful downstream constraints even when the available observational data is substantially reduced.


\end{enumerate}

Together, these experiments motivate a shift from the use of foundation models only as general-purpose feature extractors toward the use of their learned representations as \emph{discovery spaces}. 
The algorithm developed in this work is especially suitable for wide-field surveys like Roman, Euclid, and Rubin/LSST, which provide large datasets with uniform coverage across wide sky areas.



This paper presents the first result of our \textit{Learning JWST}
series, which explores the application of foundation models and deep-learning
methods to {\em JWST} imaging and spectroscopic data. \textit{Learning JWST} II
and III will extend this framework to slitless-spectroscopic {\em JWST} surveys,
with a focus on automatic and reproducible emission-line galaxy searches in ASPIRE \citep{Wang2023ASPIRE} and EIGER \citep{Kashino2023EIGER}.

The code and data will be available online when the paper is accepted. These include the training set, the literature-compiled LRD list, the high-redshift galaxies and LRD benchmark samples, and the rkNN-selected LRD candidates. 

\begin{acknowledgments}
This work is based on observations made with the NASA/ESA/CSA James Webb Space Telescope. The data were obtained from the Mikulski Archive for Space Telescopes at the Space Telescope Science Institute, which is operated by the Association of Universities for Research in Astronomy, Inc., under NASA contract NAS 5-03127 for JWST.

Y.Z. acknowledge support from the NIRCam Science Team contract to the
University of Arizona, NAS5-02105.

This research made use of astrodendro, a Python package to compute dendrograms of Astronomical data (http://www.dendrograms.org/)

This manuscript is created with assistance from AI tools. Specifically, we use ChatGPT and Claude to revise and proofread the manuscript. We have reviewed and edited this work and have retained records of our generative AI uses. We also use ChatGPT \citep{openai_chatgpt_2026} and Claude \citep{anthropic_claude_2026} to assist coding. The authors are responsible for all scientific content.


\end{acknowledgments}

\software{PyTorch, NumPy, pandas, scikit-learn, UMAP \citep{McInnes2018},
Astropy, EAZY \citep{Brammer2008}, Astrodendro, TikZ, Chatgpt, Claude}

\bibliography{references}
\bibliographystyle{aasjournalv7}

\end{document}